\documentstyle[floats,twocolumn,aps,psfig,prl]{revtex}

\begin{document}
\draft

%***********    This is for two columns 
\twocolumn[\hsize\textwidth\columnwidth\hsize\csname 
@twocolumnfalse\endcsname

%Title of paper

\title{Pressure dependence of the magnetization in the ferromagnetic 
superconductor UGe$_2$}

\author{C. Pfleiderer$^1$ and A. D. Huxley$^2$}

\address{
$^1$Physikalisches Institut, Universit\"at Karlsruhe, Wolfgang-Gaede-Str. 1,  D-76128 Karlsruhe, Germany\\
$^2$D\'{e}partment de Recherche Fondamentale sur la Mati\`{e}re 
Condens\'{e}e, SPSMS, CEA Grenoble, 38054 Grenoble Cedex 9, France\\
}

\date{\today}
\maketitle

\begin{abstract}
The recent discovery that superconductivity occurs in several clean itinerant ferromagnets close to low temperature magnetic instabilities naturally invites an interpretation based on a proximity to quantum criticality.
Here we report measurements of the pressure dependence of the low 
temperature magnetisation in one of these materials, UGe$_2$. Our results 
show that both of the magnetic transitions observed in this material as a 
function of pressure are first order transitions and do not therefore correspond to quantum critical points. Further we find that the known pressure dependence of the superconducting transition is not reflected in the pressure 
dependence of the static susceptibility. This demonstrates that the 
spectrum of excitations giving superconductivity is not that normally 
associated with a proximity to quantum criticality in weak itinerant 
ferromagnets. In contrast our data suggest that instead the pairing spectrum
might be related to a sharp spike in the electronic density of states that also drives one of the magnetic transitions.
\end{abstract}

% insert suggested PACS numbers in braces on next line
\vspace{-0.8cm}
% body of paper here
% ***********    This is for two columns 
\vskip2pc]

%\subsection{Introduction}
The possible co-existence of superconductivity and ferromagnetism, 
although considered as a theoretical possibility for idealised weak 
itinerant ferromagnets over 20 years ago \cite{gin57,fay80} has only 
recently been demonstrated to occur experimentally 
\cite{sax00,pfl01,aok01}. The theoretical calculations assumed the 
superconductivity to be mediated by an abundance of low-energy 
small-wavevector magnetic excitations. These excitations become 
prevalent near a ferromagnetic quantum critical point (QCP),
that is at the value of the pressure (or another control parameter) at 
which a second order transition is driven to zero temperature 
and at which 
the longitudinal magnetic susceptibility becomes singular. More recent 
theoretical work suggests that in an isotropic material a coupling 
between transverse and longitudinal excitations, which is present 
only in the ferromagnetic phase, should give a much higher 
superconducting transition in the ferromagnetic state 
\cite{kir01}. The presence of crystalline anisotropy has also been 
considered, and was shown to circumvent the depression of the 
superconducting critical temperature exactly at the QCP itself \cite{rou01}. 

%uge2
For UGe$_2$ it has already been established that in the limit of zero 
temperature the transition from ferromagnetism to paramagnetism as the 
pressure is increased through $p_{c} \approx 15.8$\,kbar is first order 
\cite{hux99}. This transition therefore does not correspond to a QCP. 
However, at lower pressures the temperature dependence of the magnetisation 
shows a sharp change at a pressure dependent temperature $T_{x}(p)$ well 
below the Curie temperature. $T_x$ decreases with $p$ and vanishes at $p_{x} 
\approx 12.2$\,kbar. The 
superconducting transition temperature, $T_{s}$, and superconducting coupling 
parameter are largest at pressures close to $p_{x}$ \cite{she01}. This 
would be naturally explained in the spirit of the above theory if $T_x$ were 
to correspond to a second order transition, with $p_{x}$ a QCP for this 
transition. A detailed explanation along these lines has indeed been proposed 
\cite{mia01} in which $T_{x}$ is identified with the formation of a 
simultaneous charge and spin density wave (CSDW). Theoretically the 
formation of a CSDW would lead to a change in the temperature 
evolution of the magnetic moment, as well as an enhancement of the 
longitudinal magnetic susceptibility \cite{mia01} similar to that 
near to a simple ferromagnetic QCP. Although band structure calculations 
\cite{pic01,yam01} indicate that a spin-majority Fermi-surface sheet 
could become nested as a function of the magnetic polarisation, a 
necessary condition for a CSDW to arise, extensive neutron diffraction 
studies \cite{flo01} have as yet failed to detect any static order due to a 
CSDW. 

% summary
In this Letter we establish for the first time that the low $T$ ordered 
moment 
(i.e. the ferromagnetic order parameter) and therefore a first order 
derivative of the free energy changes abruptly at $p_{x}$. 
Thus
there is unambiguously a first-order transition between two ferromagnetic 
phases at $p_{x}$ and therefore no QCP. We will refer to the high pressure 
phase as FM1 and the low pressure phase as FM2. 

% summary continued
Although the low-field low-temperature uniform 
longitudinal susceptibility undergoes a large change between FM1 and 
FM2, we show that it is almost pressure independent within each phase and is 
thus not correlated with $T_{s}(p)$ far away from $p_{x}$. Above $p_{x}$ 
the transition FM1$\to$ FM2 can be induced by a magnetic field. We 
find that the field at which the transition occurs, $H_{x}$, depends 
on $p$ but the magnetic polarisation at $H_x$ is only weakly $p$ 
dependent. This shows that the FM1$\to$FM2 transition occurs at a 
particular spin 
splitting between the majority and minority spin bands as 
would occur when the Fermi energy passes through a sharp maximum 
in the electronic density of states for one spin direction. If virtual 
excitations to states at this maximum were also associated with the 
superconducting pairing 
mechanism a pairing spectrum peaked at finite energy (in the extreme limit an 
Einstein spectrum) would result. 
We show that this provides a natural relationship between $T_s$ in 
zero field and the field necessary to induce the transition between the two 
magnetic phases for $p>p_x$. Thus the
pressure dependence of $T_{s}$, which was the motivation for previously 
supposing that
there was a QCP at $p_{x}$, can be explained without 
invoking a QCP.

% fig 1 mu(T)
\begin{figure}
\centerline{\psfig{file=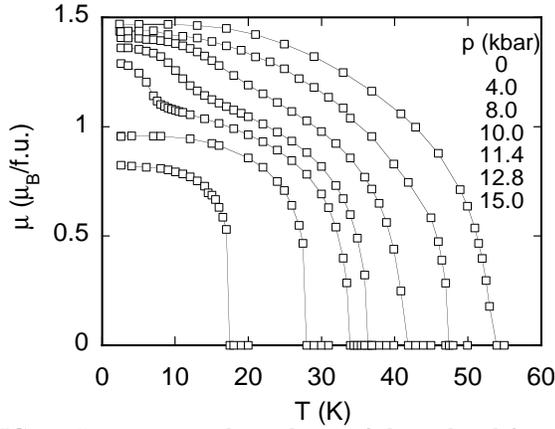,clip=,width=7.5cm}}
\caption{Temperature dependence of the ordered ferromagnetic moment, 
$\mu$, in the limit of zero field, deduced in the usual way from 
measured hysteresis loops. Curves correspond from top to bottom to 
the pressures indicated in the top right corner of the frame. The error bars are 
much smaller than the symbols.
}
\vspace{-0.5cm}
\end{figure}

% expt details
Two different single crystals cut by spark erosion from larger crystals grown 
by the Czochralski technique were studied. The larger was a cylinder of 
diameter 2.4\,mm and length 5\,mm parallel to the easy magnetic $a$-axis, 
while the smaller was a plate also parallel to this axis (glued to a small 
washer to fix its orientation in the pressure cell). Other parts of the 
larger crystal had previously been studied and found to have residual 
resistivity ratios of order 100 (current parallel to the $b$ axis) 
\cite{sax00,hux01}. The larger sample was also confirmed to become 
superconducting under pressure in a separate a.c. susceptibility measurement 
\cite{vol01b}. Here we do not distinguish further between the two samples 
since they gave equivalent results, with only 
small differences in the widths of the various transitions. The d.c. 
magnetization was measured with a non-magnetic Cu:Be clamp cell using a 
methanol:ethanol (1:4) pressure transmitting medium in a commercial vibrating 
sample magnetometer (VSM). The pressure was determined from the 
superconducting 
transition of Sn. The empty pressure cell generated a very weakly $T$ 
and $H$ dependent background contribution that was smaller than two 
percent of the signal from the larger sample in the ferromagnetic state at 
low temperature. The data shown have been corrected for this 
background. The samples were seen to float freely within the 
pressure medium before and after the experiment, while their 
orientation was constrained by the bore of the teflon sealing capsule 
in the cell (internal diameter 2.5\,mm, external diameter 3.0\,mm). 
Systematic errors in the overall calibration of the magnetometer 
when using the pressure cell mean that the absolute accuracy in determining 
the 
magnetisation is only about 5 percent. We have therefore scaled the 
data for each sample by a constant factor close to unity to give the correct 
ordered moment of 1.5 $\mu_{B}$/formula at zero pressure. The
experimental error in measuring the relative changes of magnetisation 
is in contrast much smaller, and smaller than the size of the data points 
used in the various figures.

\begin{figure}[t]
\centerline{\psfig{file=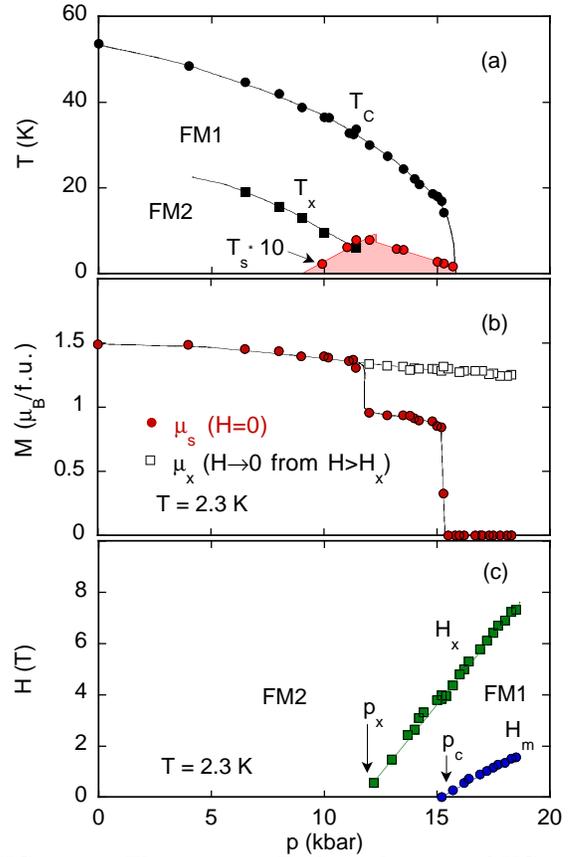,clip=,width=7.5cm}}
\caption{(a) The $p$ versus $T$ phase diagram of UGe$_2$. $T_{C}$ is 
the Curie temperature and $T_x$ is defined in the text. 
$T_{s}$ is the superconducting temperature (onset) from ref. 14. The 
lines through the data points are a guide to the eye, noting that 
$T_{s}$ might change discontinuously at $p_{x}$ and $p_{c}$. (b) The 
pressure dependence of $\mu$ in zero field at 2.3\,K, (full circles). 
The moment obtained by extrapolating the data from above $H_{x}$ to 
zero field (squares) is also shown when this is different. (c) the 
pressure evolution of the fields $H_x$ and $H_m$ of metamagnetic 
transitions (at which $dM/dH$ has a local maximum) at 2.3\,K.
}
\vspace{-0.5cm}
\end{figure}
%

% results : moment vs T and construction of phase diagram
In Fig. 1 the temperature dependence of the ordered magnetic moment, 
$\mu(T)$, in zero field at different pressures is shown (we use the 
symbol $\mu$ for the ordered moment extrapolated to zero field and 
$M$ more generally for the magnetisation in a field). A clear change in 
the $T$ dependence of $\mu(T)$ occurs at $T_x$, as has been 
previously reported \cite{hux01,tat01b}, where $T_x$ decreases with 
increasing $p$ and disappears as $p \to p_x \approx 12.2$\,kbar. 
While $T_{C}$ can be conveniently defined as the point where $-dM/dT$ 
is a maximum in a small field (we used 0.02 T) the determination of 
$T_{x}$ is slightly more subjective and is taken as the position of a 
local maximum of $d\mu/dT$. The resulting $p$ versus $T$ phase 
diagram for UGe$_2$ constructed from the present measurements is 
shown in Fig. 2(a) along with $T_s(p)$ taken from reference 
\cite{hux01}. $T_{x}$ cannot be assigned from the present 
magnetisation measurements below $\sim$\,6\,kbar. The position of a 
peak in the temperature derivative of the resistivity reported by Oomi 
et al. \cite{oom95} can however be used to extend the $T_{x}$ line to 
give $T_{x} \approx 30$\,K at $p=0$. In the following we focus on the 
pressure dependence of the magnetisation at low $T$.

% Main result moment at T=0  vs pressure
The $p$ dependence of the low $T$ ordered moment $\mu$ at 2.3\,K is 
shown in Fig. 2(b). Striking features are the abrupt changes of 
$\mu(p)$ on crossing $p_x$ and $p_c$, respectively. This is the main 
new result. It shows that the transition from FM2 $\to$ FM1 at 
$p_x$ is a first-order transition in the limit of $T=0$, and confirms 
that the transition from the ferromagnetic state FM1 to the 
paramagnetic phase at $p_{c}$ is also first order 
\cite{hux99,hux01,ter01}. 

% field induced transition - overview
The field dependence of the magnetisation at 2.3 K for different $p$ 
is shown in Fig. 3. For pressure $p>p_x$ a large increase of 
nearly 50\,\% in the magnetisation is observed at a field $H_{x}$, 
($H_{x}$, defined as the field at which $dM/dH$ has a local maximum is 
plotted as a function of pressure in Fig. 2(c)). For $p>p_c$ the magnetisation
undergoes a second increase 
at a lower field $H_m$ corresponding to the transition from the 
paramagnetic phase to FM1. Interestingly, the uniform susceptibility 
given by the slope dM/dH has almost constant values independent of 
the pressure within each 
phase; in the FM1 phase it is greater than in the FM2 phase 
but less than in the paramagnetic state above $p_c$

% fig 3
\begin{figure}
\centerline{\psfig{file=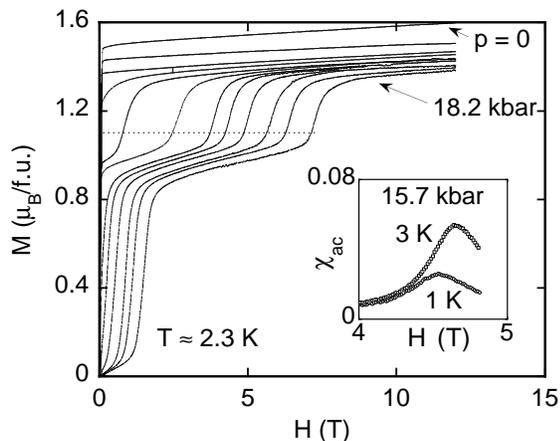,clip=,width=7.5cm}}
\caption{The field dependence of the easy-axis magnetization at 
$2.3$\,K for various pressures. The broken line passes through the 
points $H_x$ at which $dM/dH$ has a local maximum. The magnetisation 
at $H_{x}$ is almost independent of pressure and suggests that the 
transition FM1$\to$FM2 occurs at a fixed value for the splitting 
between spin majority and minority bands. Curves correspond from top to bottom to $p=0, 6.5, 9.0, 11.1, 12.8, 13.8, 15.3, 15.5, 16.0, 16.7, 17.3$ and 18.2\,kbar. The inset shows the a.c. susceptibility in S.I. units measured as a function of field at 15.7\,kbar at 1\,K and 3\,K.}
\vspace{-0.5cm}
\end{figure}

% discussion of finite field transitions - 1st order
The existence of metamagnetic behavior just above $p_x$ is in itself an indication that the transition between the two magnetic phases is first order.
We now consider further the transitions at $H_x$. Hysteresis 
loops of the d.c. magnetisation in low fields show that the sample 
is already mono-domain in a field of 0.02 Tesla and therefore no 
hysteresis would normally be expected at much higher fields of several Tesla. 
However we observe hysteresis  of a few mT (not visible on the scale of Fig. 3) at both $H_m$ and $H_x$ in careful measurements. The evidence for such hysteresis at $H_x$ and $H_m$ is demonstrated unambiguously by comparing the present data to measurements of the a.c. susceptibility, $\chi_{ac}$.
In the inset of Fig. 3, $\chi_{ac}$ is shown as a function of field in the 
vicinity of $H_{x}$ at a pressure of 15.7 kbar (from reference \cite{hux01}). 
The amplitude of the peak in the a.c. susceptibility at 3\,K is smaller than the derivative of the uniform magnetisation $dM/dH$ at $H_{x}$, despite the fact that the peak in the a.c. measurement is slightly sharper than the d.c. transition width. Further, $dM/dH$ at H$_{x}$, decreases with increasing $T$, whereas the amplitude of the peak in $\chi_{a.c.}$ increases with $T$ (at least up to 5\,K). This shows that the a.c. measurement traces minor hysteresis loops in the vicinity of $H_{x}$ that become wider at lower $T$. The same result is also found for the transition at $H_{m}$ \cite{hux99}. The observation of 
hysteresis supports our previous conclusion that the transition between the FM1 and FM2 phases is first order at low temperature; for a first order transition 
a phase can exist metastably in a limited region beyond that in which it is thermodynamically stable.

%expt discussion of the width
%To complete the presentation of the experimental results we briefly 
%consider the possible inconsistency of the widths of the transitions 
%in Fig. 3 with first order behaviour. Measurements of the thermal 
%expansion \cite{oom93,nis94} provide a relationship between a change 
%in volume, $\omega_S=\Delta V/V$, and the ordered magnetic moment of 
%UGe$_{2}$; $\omega_S=7\cdot 10^{-4} (\mu/\mu_B)^2$. Since we have 
%used as large a crystal as possible inside the teflon capsule to 
%maximize the measured signal, leaving a small amount of pressure 
%liquid between the circumference of the sample and the teflon 
%capsule, our measurements could in an extreme case constrain the 
%sample volume to be constant. The increase in pressure as the 
%transition at $H_{x}$ is crossed at constant volume is then 
%calculated to be $\Delta p \approx \Delta\omega_S B_{s}$ where 
%$B_{s}$ is the bulk modulus of the sample (estimated from the phonon 
%velocities \cite{ray02} to be of the order $B_{s} \approx 
%2000$\,kbar). This is of the required magnitude to explain the 
%observed widths of the transitions showing a technical origin and 
%there is thus no inconsistency with first order behaviour.

% fluctuations ferro? % our model
We now discuss the maximum of $T_s(p)$ near $p_x$, which was previously 
the main motivation to suppose that $p_x$ marked a QCP. We focus on the 
FM1 phase (i.e. $p_{x}<p<p_{c}$) where the superconducting 
transitions are much sharper. For ferromagnetically mediated pairing 
$T_{s}$ can be estimated as $T_{s}=\theta \,e^{-\gamma/g\Delta\gamma}$, where 
$\Delta\gamma$ is that part of the linear temperature dependence of 
the normal state electronic heat capacity, $\gamma$, associated with 
the excitations responsible for pairing \cite{mil88,gil97}. $\theta$ 
is the characteristic energy of these excitations and g the 
effectiveness of this pairing channel (we consider the 
superconductivity to be non s-wave with $g < 1$ and constant). In the 
usual description of itinerant ferromagnetism the spectrum of 
longitudinal magnetic excitations is assumed to be a Lorentzian 
peaked at zero energy ($\omega$) and wavevector transfer ($q$) 
\cite{gil97}. For such a spectrum and conventional $q$- and 
$\omega$-independent mode-mode coupling $\Delta\gamma$ is directly 
related to the $T$ dependence of $\mu^2$ at low $T$. Our experiment 
shows that the temperature dependence is much weaker for pressures just above 
$p_{x}$ than just below $p_c$ (Fig. 1). $\Delta\gamma$ is therefore expected 
to increase significantly with $p$ even though the static 
longitudinal susceptibility defined as $dM/dH$ (Fig. 3) is 
experimentally almost independent of $p$ between $p_{x}$ and $p_{c}$. 
The latter point could still be reconciled with a Lorentzian spectrum 
if the width of the spectrum increases either in $q$ or $\omega$. 
However, experimentally $\gamma$ is known to be almost constant 
between $p_{x}$ and $p_{c}$ \cite{tat01} and thus $T_{s}(p)$ would 
also increase with $p$ if superconductivity was indeed due to a 
Lorentzian spectrum of excitations. This is in stark contrast with 
the observed decrease of $T_{s}$ with p. Thus a simple spectrum of 
longitudinal magnetic excitations of the type usually considered
near a ferromagnetic QCP cannot account for our experimental observations.

%Our model
In the following we outline a mechanism that qualitatively explains the 
observed pressure dependence of $T_s$ consistently with first order 
transitions at $p_x$ and $H_x$. The mechanism is based on our observation that the FM1$\to$FM2 transition occurs at a constant magnetisation independent of the pressure. This strongly suggests that the transition takes place when the 
Fermi-energy crosses a sharp maximum in the electronic density of states (DOS) for one spin-polarisation. In the FM1 phase an applied field parallel to the easy magnetic axis leads to an additional Zeeman splitting between the majority and minority spin bands, which drives the Fermi-energy through this maximum. $\mu_{B}H_{x}$ is then proportional to the energy of the maximum in the DOS relative to the Fermi-energy in zero field. If we suppose that the superconducting pairing involves virtual excitations that access the same feature in the DOS the pairing strength and therefore $\Delta\gamma$ decrease strongly as the feature becomes more remote from the Fermi-surface. Thus, for example, the decrease of $T_{s}$ with p in the FM1 phase is a naturally
linked with the increase in $H_{x}$. Further support that the 
excitations responsible for pairing have a spectrum peaked at a 
finite energy proportional to $H_{x}$ comes from the measured upper 
critical field for fields along the c-axis (i.e. perpendicular to 
the easy axis). It has previously been shown that the temperature dependence of the upper critical field is well modelled by a strong coupling calculation 
assuming an Einstein spectrum for the pairing interaction \cite{she01}; the 
position of the peak in the spectrum obtained by fitting the measured 
upper critical field to this model increases with pressure in the FM1 phase as we have described.   

%FS
Spectroscopic measurements capable of detecting a sharp peak in the DOS 
have not yet been reported. Quantum oscillation measurements as a function of $p$ were however recently published \cite{ter01} and so we briefly examine whether these can be reconciled with a sharp peak in the D.O.S. The striking feature in the quantum oscillation data is that the electronic masses of all  
the detected orbits are much higher in the FM1 phase than in the FM2 phase (some frequencies remain similar while others differ substantially). Large mass renormalisations in heavy fermion materials are usually attributed to a Kondo-like mechanism where narrow f-electron bands in the pure ordered 
system play the roles of the Kondo impurity states lying just below the Fermi energy in the original Kondo analysis \cite{kon69}. Assuming a similar mechanism is responsible for the large effective masses observed in the FM1 phase of UGe$_2$ the much smaller masses in the FM2 phase require a destruction of the mechanism. Such a destruction would indeed occur if the Fermi level were to cross one of the narrow bands responsible for the resonant mass enhancement. 

%conclusion
To conclude, we note that understanding the emergence of new physical behaviours close to quantum criticality represents one of the central themes in contemporary studies of correlated electron physics. The case of a ferromagnetic QCP is particularly important since the order parameter is directly measurable by macroscopic techniques. However we have shown that although superconductivity in UGe$_{2}$ is intimately related to a proximity to a magnetic phase transition there is no quantum criticality associated with the suppression of this transition to zero temperature at pressure $p_{x}$. The implication is that new ground states (in this case non-conventional superconductivity) can emerge in strongly correlated electron systems due to a much wider range of circumstances than has hitherto been supposed.

Discussions with J. Flouquet, H. v. L\"ohneysen, G. G. Lonzarich, V. 
Mineev, K. Miyake,  E. Ressouche, A. Rosch, J. P. Sanchez,  M. Uhlarz 
and R. Vollmer are gratefully acknowledged. We also acknowledge financial 
support from the 
Deutsche Forschungsgemeinschaft, CEA - Direction des Sciences de la 
Mati\`{e}re, and the European Science Foundation FERLIN program.

\end{document}